# The Appliance Pervasive of Internet of Things in Healthcare Systems

**Mir Sajjad Hussain Talpur**
**School of Information Science & Engineering,**
**Central South University (CSU), 410083 - Changsha, China.**
mirsajjadhussain@gmail.com

**Abstract**

In fact, information systems are the foundation of new productivity sources, medical organizational forms, and erection of a global economy. IoT based healthcare systems play a significant role in ICT and have contribution in growth of medical information systems, which are underpinning of recent medical and economic development strategies. However, to take advantages of IoT, it is essential that medical enterprises and community should trust the IoT systems in terms of performance, security, privacy, reliability and return-on-investment, which are open challenges of current IoT systems. For heightening of healthcare system; tracking, tracing and monitoring of patients and medical objects are more essential. But due to the inadequate healthcare situation, medical environment, medical technologies and the unique requirements of some healthcare applications, the obtainable tools cannot meet them accurately. The tracking, tracing and monitoring of patients and healthcare actors activities in healthcare system are challenging research directions for IoT researchers. State-of-the-art IoT based healthcare system should be developed which ensure the safety of patients and other healthcare activities. With this manuscript, we elaborate the essential role of IoT in healthcare systems; immense prospects of Internet of things in healthcare systems; extensive aspect of the use of IoT is dissimilar among different healthcare components and finally the participation of IoT between the useful research and present realistic applications. IoT and few other modern technologies are still in underpinning stage; mainly in the healthcare system.

## 1. Introduction

In healthcare industry, Internet of Things (IoT) provides an opportunity of discovering healthcare information about a tagged patient or medical object by browsing an Internet address or database entry that corresponds to a particular Radio-Frequency Identification (RFID) tag. But now, it is extended to the general idea of medical things, especially healthcare everyday objects, those are readable, recognizable, locatable, addressable, and/or controllable. These objects may be equipped with devices such as sensors, actuators, and RFID tags, in order to allow patients, doctors, equipments and other healthcare actors to be connected anytime and anywhere with anything and anyone.

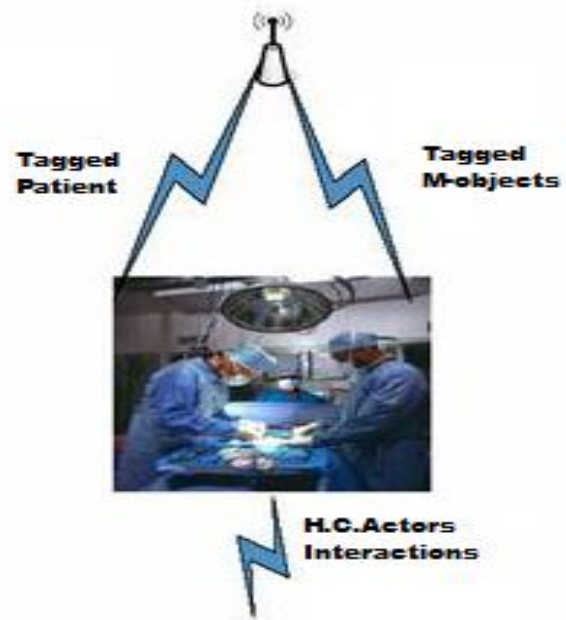

Fig1. IoT modeled Healthcare Actors Interactions

Through the medical object-to-person and object-to-object communications, IoT enables a wide range of smart applications and services to cope with many of the challenges that individuals and organizations face in their everyday lives, such as smart healthcare, smart home, smart earth, smart kitchen, smart transportation and smart office, etc. IoT is not a simple extension of the Internet or an aggregation of Internet systems, and it covers a wide range of technologies including tagging, sensing, networking, computing, storage, and control, which together build feasible complex cybernetic physical and social systems to support these smart applications [1], [2]. Internet of things has dynamic capabilities to connect D2M (Device-to-Machine), O2O (Object-to-Object), P2D (Patient-to-Doctor), P2M (Patient-to-Machine), D2M (Doctor-to-Machine), S2M (Sensor-to-Mobile), M2H (Mobile-to-Human), T2R(Tag-to-Reader), intelligently connects humans, machines, smart devices, and dynamic systems which ensure the effective healthcare system, health monitoring system, medical assets monitoring and medical waste management system[3].

Medical has ever remained one of the major applications of internet. The collaboration of internet and medical formed a sub-field e-health. An application of Internet and other related technologies in healthcare industry to improve the access, efficiency, effectiveness, quality of clinical and business processes utilized by healthcare organizations, practitioners, patients in an effort to improve the health status of patients [4]. Along with many other services, online appointment services in particular are the most common e-health services [3]. E-health is one of small component of IoT based healthcare management system, in which an online interaction of a patient to a doctor is made possible and easy along with easy access to online healthcare record checking for the patients. Whereas IoT based healthcare system consists all this plus identification and tracking of patients and doctors locations, tracking of patient's health records and tracking locations of hospital equipment etc. IoT have also enabled intelligent behavior of some equipment which alarm automatically when near to expiry or auto informing a relevant doctor if concerned patient is brought to Intensive Care Unit. *"Patients and medical objects tracking, tracing and monitoring are demanded by world wide healthcare institutions, governments and other organizations, the universal expectation of all healthcare organizations wants state-of-the-art patients tracking system, has put endorse superior requirements"* [5]. In order to enhance the protection of patients, improve the healthcare system and make them competitiveness of patient care, must develop an effective patient traceability system, traceability system of patients care should have dynamic and efficient management of the healthcare system [6]. Therefore, it has immense challenge for researcher or developer how to use information and communication technology; so the implementation of healthcare management system has become one of key areas. The IoT is a key technology that is quickly gaining ground in the development of modern secure wireless communications.

| *Tracking/ Tracing* | *Sensing* | *Identification and authentication* | *Real time data collection* |
|---|---|---|---|
| Patients tracking and tracing at hospitals and outside to monitor the patient flow | Intelligent medication monitoring (pregnant women or elderly) at home or hospital | Protecting patient privacy | Automatic data collection and transfer is mostly aimed at reducing form processing time, process automation. |
| Tracking of patient location | home | Patient identification to reduce harmful incidents | Automated care and procedure auditing, and medical information management. |
| Tracking of drugs, supplies and procedures performed on patient | Sensing will be able to do real time monitoring of patients. Parameters such as blood pressure, glucose levels, heart and breathing rate | Eliminate wrong patient/wrong surgery | Relates to integrating IOT, RFID technology with other health information and clinical application technologies. |
| Accounting patient time in emergency department | Sensor devices enable function centered on patients, and in particular on diagnosing patient conditions, providing real-time information on patient health indicators | Patient identification to avoid wrong drug, dose, time, procedure | Integrating state-of-the-art physiological parameters monitoring time interval in order to determine actual time period. |

Table.1. Potential of IoT in Healthcare System

The main idea of this thought is the pervasive existence around us of a mixture of stuff or substance; such as

Radio-Frequency Identification (RFID) tags, EPC technology, tiny sensors, actuators, smart phones, and so on. These things are capable to interact with each other and collaborate with their neighbors through the unique addressing schemes, in order to achieve their goals [7, 8]. In healthcare industry, IoT is applied for patient tracking by offering intelligent jackets, wristbands containing RFID tags. The tags interact with hospital information system for automating administrative tasks like patient admissions, patient transfers and discharges. *"The U.S Food and Drug Administration (FDA) have recently approved a tag called "veri-chip" in humans. These tags facilitate disoriented, elderly patients more safe by storing a detailed health information record. With this FDA approval in the headlines, suppliers have begun to offer patient wristbands containing tags"* [3]. Even though IoT`s application in patients tracking is less talked about, serious situations could be avoided implementing RFID tags to specific patients atleast, think the severity of this incident that occurred in phoenix recently to emphasize the importance in use of RFID tags, when a patient with dementia wandered from his / her room was found in the storage area, after 3 consecutive days.

## 2. The Structure of IoT

Through the Internet of things, anything in the healthcare system can be identified tracked and monitored on demand anytime anywhere [1]. Internet of things is considered as remarkable revolution after the blooming of Internet with ICT based industry. Internet of things has three basic components, namely RFID systems, middleware systems and Internet systems Savant. RFID system is one of the major components of IOT and it enables data to be transmitted by a portable device, called *"a tag"*, which is read by an RFID reader and processed according to the needs of a particular application [9]. The data transmitted by the tag may provide identification or location information, or specifics about the patient tagged, such as (e.g. patient ID, age, sex, blood pressure, glucose level) therefore the RFID systems can be used to monitor healthcare objects in real-time, without the need of being in line-of-sight. This allows mapping of real world healthcare system into the virtual world system. Middleware savant system is software that bridges RFID hardware and healthcare applications [10].

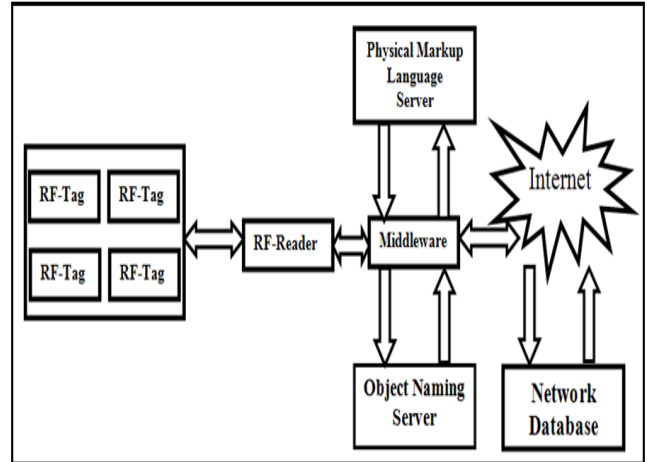

Fig 2. Basic composition diagram of IoT

Indisputably, the primary means of medical data gathering for any Radio frequency identification deployment and it consists of savant server, Object naming service servers, Physical Markup Language server and the corresponding medical data server software [11]. Internet system consists of state-of-the-art computer systems and secure network servers as shown figure2.Vision of healthcare technology mainly rely on accurate patient recognition in reducing healthcare complications, errors and harmful drug effects. The emboldens of IoT technologies in healthcare will ensure the healthcare safety, exact patient, accurate drug, proper dose, right way and exact time, by complying with the principles and regulations of HIPAA (Health Insurance Portability and Accountability Act of 1996) which mentions the standards of data exchange with protection and confidentiality of patient information, JCAHO (Joint Commission on Accreditation of Healthcare Organizations that emphasizes positive patient identification) and AHA (American Hospital Association) stressing guidelines for tamper proof non-transferable wristband minimizing the risk of losing transferred data. The Internet of Things(IoT) system in patient tracking provides non-transferable accurate patient identification which will ensure the safety of patient and reduced the harmful incidents in healthcare system, The IoT will improve efficiency of healthcare system, introduces affordable low cost RFID tags and tiny sensors which are associated to medical devices to reduce service costs, improve the quality of service, control the medical defects, improve identification and authentication of patients, automatic data collection and sensing system [12].IoT has vision to swagger the healthcare based smart communication technologies in order to connect the healthcare actors anywhere and anytime. That is why; Internet of Things played an exigent role in healthcare system.

## 3. Functioning Philosophy of IoT System

Internet of Things is a technological revolution that represents the future of computing and communications, and its development depends on dynamic technical innovation in a number of important fields, from wireless sensors to nanotechnology [1]. The essential functioning principles of Internet of things based on Radio Frequency Identification which is known as the soul of IoT, EPC technology used global unified products coding and wireless communications technology in order to tracing of healthcare objects, swank integrity of healthcare system [13,14]. Medical products, labeled with EPC code stored electronic tags. Furthermore, in the whole life cycle of the medical product; the EPC code easily recognizes a medical product, as EPC codes for the index in real-time query and modify medical objects information from the healthcare network, but also use it as signs, all movement in the healthcare system to find the medical product tracking. Entire healthcare actors are connected with RFID tags, when a RFID reader read range in its tags to monitor the existence, the label contained in the EPC and its linked data transfer to the savant middleware [10]. Initially the electronic medical product code data is key; in the local Object naming server (ONS) contains the information of medical product for EPC information server's network address, and savant query. According to the EPC information server address, access to medical commodity specific information, the essential cure, to transmit the information back-end healthcare applications to do a deeper level of computing ,at the same time, local EPC information server and source of this information server to record the reader to read and modify the corresponding information.

## 4. Networking Structure of Healthcare System

### A. Healthcare Activities Analysis on the basis of IoT

Healthcare system is the organization of citizens, medical institutions, and healthcare resources to deliver healthcare services to meet the health needs of citizens [8].The healthcare system has been overwhelmed by problems such as patient diagnoses being written illegitimately on paper, doctors not being able to easily access patient information, and limitations on time, space, and personnel for monitoring patients. With advancements in technology, opportunities exist to improve the current state of healthcare to minimize some of these problems and provide more personalized service. Because characteristics of healthcare system, information flow is necessary to pass a higher, in time, need to be quicker, in space, and requires more strict storage conditions. Therefore, it made healthcare traceability an advanced technical requirement, while the development and stability of the healthcare system made a greater requirement [11].All nodes in healthcare system should be a integrated, harmonious division of medical staff and cooperation, and layout of the traditional healthcare system is comparatively complex, healthcare actors activities relatively unstable and so the delay caused by the healthcare organizations and asymmetric information, splits the chain between healthcare actors and enterprise [13]. It can be seen that the traditional healthcare system are often in motion or loose state of the information, timeliness, accuracy, and sharing, based on EPC technology, the application of medical things; a good solution to the above problem. An EPC have tag read and write data function, easy compactness and diversification of the shape, reusable, good penetration and data capacity and other characteristics, can adapt to frequent changes in the healthcare information system, communicate data, acquisition in a timely manner and system commands, widely used in medical products warehouse management, hospital transportation management, medical production management. Healthcare actors tracking, identification and significantly reduces the health complications [13]. Therefore, IoT application for healthcare, the use of RFID and secure databases will integrate all types of medical production, movement and the effective information quality and protection [15]. Medical information collection, storage, transmission, the healthcare system and safety of healthcare actors combined through the entire system, and established healthcare actors traceable system based on IoT. Which archive a summary of the medical data and information in the entire healthcare system to make regulatory and patient information track in the platform, and guarantee transparency throughout the healthcare system process [16].

### B. Brunt of Internet of things on Healthcare System

The brunt of patient identification and medical objects identification processes in healthcare system for instance: patient`s identification to reduce harmful incidents to patients (e.g. wrong drug/dose/time etc). Relation to staff, identification and authentication are most frequently used to grant access and to improve employee morale by addressing patient safety issues. In relation to assets, identification and authentication is predominantly used to meet the requirements of security procedures, to avoid thefts or losses of important instruments and products, so in tracking and identification of patients, healthcare staff identification and medical assets identification. In the entire system, the main objective of IoT is to manage the identification process of healthcare actors [8]. RFID tags and tiny sensors are attached to patients and patients also wear bracelet and it has unique identification number of RFID tag, which process and record all relevant information collected on the birth, at the same time, applied RFID chip, tiny sensor in order to manage patients' medical information, assets and other healthcare actors' information.

## 5. Intensification Necessitate in Health Monitoring System for Quality of Service

Internet of things provides an effective way of real time remote monitoring system of healthcare actors through RFID tags, sensors, and actuators. The RFID tags in healthcare may be applied to patients, assets, medical staff and other objects, allowing the readers on gate frames, hospital wards and other treatment areas of hospital to detect and record interactions. IoT is applied for patient tracking by offering wristbands containing RFID tags. The tags interact with healthcare information system for automating managerial everyday tasks like patients' admissions, transfers and discharges [17], [18]. IoT applications in medical administration to streamline the processes and reduce healthcare harmful incidents the above arguments prove that Internet of things ensure the safety of patients and quality of service.

## 6. Conclusion

The name, Internet of Things, is syntactically composed of two terms, Internet and Things. As a result, it's usually considered that there are two versions of IoT, "Internet oriented" or "Things oriented". The "Things oriented" version focus on the technology developed to improve object visibility, such as awareness of its status, current location. This is undoubtedly a key component of the path to the full deployment of the IoT vision but it's not only one [19]. In past decades, the Internet has connected numerous devices. Why don't we take advantage of the existing Internet technology and connect smart objects around the world? [13]. We discussed the significance of IoT especially in healthcare system; immense prospects of Internet of things in HCS; extensive aspect of the use of IoT is dissimilar among different healthcare components and finally the participation of IoT between the useful research and present realistic applications. IoT and few other advance technologies are still in underpinning stage; mainly in the healthcare system. This article tries to emphasize a healthcare system not only to realize the illustration and traceability of healthcare actors guarantee the quality but also effectively controls healthcare actors.

**Acknowledgment**

This work is supported by National Natural Science Foundation (NSF) of Peoples Republic of China under grant numbers 61073037 and 61103035, Ministry of Education fund for doctoral disciplines in Higher Education under the grant number of 20110162110043. The authors are thankful to the Ministry of Education for the financial support.

**First Author: Mir Sajjad Hussain Talpur** received master`s degree in Information Technology from Shah Abdul Latif University Khairpur mirs, Pakistan in 2003.He is currently a Doctoral Degree Candidate at the Trusted Computing Institute, School of Information Science and Engineering, Central South University, Changsha, Hunan China. He is also a Lecturer at the Information Technology Institute, Sindh Agriculture University, Sindh Pakistan. His current research interests include Internet of Things in Healthcare System.